\documentclass{PoS}
\usepackage{amsmath}
\title{Effective Field Theory and Lattice QCD approaches for hard probes in QCD matter}

\ShortTitle{EFT and Lattice QCD approaches for hard probes in QCD matter}

\author{\speaker{Miguel \'{A}ngel Escobedo}\\
        Instituto  Galego  de  F\'{i}sica  de  Altas  Enerx\'{i}as  (IGFAE),  Universidade  de  Santiago  de  Compostela,  Galicia-Spain\\
        Department of Physics, P.O. Box 35, FI-40014 University of Jyv\"askyl\"a, Finland \\
        E-mail: \email{miguelangel.escobedo@usc.es}}

\abstract{Hard Probes are an essential tool to discover the properties of the quark-gluon plasma created in heavy-ion collisions. The study of hard probes always involves taking into account very different energy scales, and this is precisely the situation in which Effective Fields Theories (EFTs) are useful. EFTs can be used to separate the short-distance and perturbative physics from the long-distance and non-perturbative one. This method combined with Lattice QCD evaluations of the long-distance effects can provide accurate and first principles results. In this proceeding, I will report recent advances in this direction. Results from an EFT computation of quarkonium $R_{AA}$ at $\sqrt{s_{NN}}=5.02$~TeV are shown for the first time here.}

\FullConference{International Conference on Hard and Electromagnetic Probes of High-Energy Nuclear Collisions\\
		30 September - 5 October 2018\\
		Aix-Les-Bains, Savoie, France}

\begin{document}

\section{Introduction}
One of the defining properties of hard probes is that the energy that is needed to create them is much larger than the thermal scales that are going to be present in the medium. Therefore, their creation process involves essentially $T=0$ physics. However, when these probes interact with the quark-gluon plasma, they do get modified. The conclusion is that comparing the behavior of the hard probes in pp and AA collisions it is possible to infer some properties of the medium created in heavy-ion collisions.

From the previous discussion, it is clear that the physics of hard probes in heavy-ion collisions always involves at least two very separated energy scales:
\begin{itemize}
\item The hard scale that is involved in their creation process.
\item All the lower energy scales related to the presence of the medium.
\end{itemize}
The presence of this hierarchy is not just an academic issue. Unexpected enhancements and suppressions appear that can make naive perturbation theory inaccurate even in cases in which the coupling constant is small. Regarding Lattice QCD, it increases the computational cost by demanding a huge difference between the lattice size and the lattice spacing. An example of this is given by the Sommerfeld factor, which represents a failure of naive perturbation theory when computing the annihilation amplitude of non-relativistic particles. 

In order to control these difficulties, it is convenient to use EFTs \cite{Weinberg:1978kz}. An EFT is a theory which reproduces the same results as another more fundamental theory (in this case QCD) but limited to low energy degrees of freedom. In other words, it agrees on the infrared with the full theory. We can construct the Lagrangian of an EFT in the following way:
\begin{itemize}
\item Identify the symmetries that are present at low energies.
\item Identify the relevant degrees of freedom.
\item Write the Lagrangian as a sum of all the operators built with the relevant degrees of freedom which respect the symmetries. The constants that multiply these operators are called Wilson coefficients. 
\item Fix the Wilson coefficients by imposing that the EFT is equivalent to the infrared of the full theory.
\end{itemize}
A last fundamental ingredient is the power counting. In general, a Lagrangian so constructed has an infinite number of terms and is not renormalizable. However, given an aimed precision it is possible to define a set of rules based on the separation of energy scales that indicate which terms in the Lagrangian need to be taken into account, restoring in this way the predictive power. In the corresponding sections, we will give examples of EFTs that are used to study hard probes.

Until now we have not discussed how to compute the long-distance effects that influence the behavior of hard probes. This can be done using Lattice QCD computations. These techniques are based on simulating QCD in a discretized Euclidean space and have been used to study many thermodynamic properties of nuclear matter, for example, the equation of state of QCD (see \cite{Borsanyi:2016ksw,Bazavov:2017dsy} for a recent computation). 

In order to realistically simulate a system using Lattice QCD one has to make sure that the size of the lattice is larger than the largest distance scale relevant to describe the system and that the lattice spacing is smaller than the smallest relevant distance scale of the system. These constraints can be too demanding when dealing with hard probes where separated energy scales are involved. The combination of Lattice QCD with EFT techniques is very interesting because short-distance contributions, which can be computed in perturbation theory, are encoded in Wilson coefficients and do not need to be simulated on the lattice. This can be used to increase the lattice spacing, reducing computational cost in this way and making possible computations that would be not otherwise. Examples of quantities relevant for the study of hard probes that can be computed non-perturbatively using this combination of techniques are the finite temperature static potential \cite{Burnier:2016mxc} and the jet broadening parameter $\hat{q}$ \cite{Panero:2013pla,Laine:2013lia,Majumder:2012sh}.

The structure of this manuscript is the following. In the next section,  we discuss heavy quarkonium and heavy quarks. We study Jet quenching in Section \ref{sec:jet}. Finally,  Section \ref{sec:concl} contains our conclusions.

\section{Heavy quarkonium and heavy quarks}
The mass of heavy quarks is much bigger than $\Lambda_{QCD}$. Then the elementary process that creates them is perturbative and, therefore, not affected by the medium. Heavy quarkonium is a bound state form by a heavy quark and an antiquark. A consequence of asymptotic freedom is that a bound state formed by quarks with a large mass $m$ is a non-relativistic system, the velocity of the heavy quarks ($v$) around the center of mass is much smaller than $1$.  In the bottom case it is estimated that in $\Upsilon(1S)$, $v^2\sim 0.1$ and the same can be said for the more deeply bound states of charmonium (In $J/\Psi$, $v^2\sim 0.3$). The non-relativistic velocity induces a separation of energy scales between the heavy quark mass and the inverse of the typical radius (that goes like $mv$) that is exploited by non-relativistic QCD (NRQCD) \cite{Caswell:1985ui,Bodwin:1994jh}.  NRQCD is an EFT that reproduces the same results as QCD for processes in which the virtuality of the particles involved is much smaller than $m^2$. It is a theory that is very commonly used to study production and decays of heavy quarkonium. 

There is an additional energy scale that is also important to study bound states. As a consequence of the heavy quarks moving with a non-relativistic momentum $mv$, the kinetic energy (and due to the virial theorem also the binding energy) is of order $mv^2$. It is possible to define an EFT that reproduces the same results as NRQCD (and therefore QCD) for processes in which the virtuality involved is smaller than $m^2v^2$. This EFT is potential NRQCD (pNRQCD) \cite{Pineda:1997bj,Brambilla:1999xf}. Its Lagrangian is the following:
\begin{align}
\label{eq:LanpNRQCD}
&\mathcal{L}_{pNRQCD}=\int\,d^{3}{\bf r}Tr\left[S^{\dagger}\left(i\partial_{0}-h_{s}\right)S+O^{\dagger}\left(iD_{0}-h_{o}\right)O\right]+V_{A}(r)Tr(O^{\dagger}{\bf r}g{\bf E}S+S^{\dagger}{\bf r}g{\bf E}O)\\ \nonumber
&+\frac{V_{B}(r)}{2}Tr(O^{\dagger}{\bf r}g{\bf E}O+O^{\dagger}O{\bf r}g{\bf E})+\mathcal{L}_{gluons}+\mathcal{L}_{q-light}\,.
\end{align}
In pNRQCD the degrees of freedom are (apart from gluons and light quarks) a color singlet field $S$ and a color octet one $O$, both of these fields contain heavy quarks and antiquarks.  The first two terms in the pNRQCD Lagrangian correspond to a Schr\"{o}dinger equation in which 
\begin{equation}
h_{s,o}=\frac{p^2}{m}+V_{s,o}\,,
\end{equation}
where $V_{s,o}$ is the singlet (octet) potential. The potential in the pNRQCD framework has a precise definition. It is a Wilson coefficient of the EFT. It is possible to use the static limit ($m\to\infty$) to match NRQCD to pNRQCD. Using this it is easy to show that $V_s$ corresponds to the static potential that can be obtained studying the Wilson loop on the lattice \cite{Brambilla:1999xf}. However, the pNRQCD framework goes beyond a potential model. Taking into account additional terms in the Lagrangian in Eq. (\ref{eq:LanpNRQCD}) we can obtain corrections that, in general, cannot be encoded in a potential. An example is the Bethe logarithm that appears when computing the spectrum at order $m\alpha_s^5$. 

In the context of quarkonium suppression in heavy-ion collisions, one of the most important questions has been whether the problem can be described using a potential model and, if so, how to define the potential. The application of non-relativistic EFTs can provide an answer to these questions. At finite temperature, on top of the energy scales related to quarkonium itself, we have to consider those induced by the medium, for example, the temperature $T$ and the Debye mass $m_D$. Integrating out these energy scales induces new versions of the non-relativistic EFTs with modified Wilson coefficients \cite{Brambilla:2008cx,Escobedo:2008sy}. For example, if we consider the temperature regime $\frac{1}{r}\gg T \gg m_D, E$ we can start with the Lagrangian in Eq. (\ref{eq:LanpNRQCD}). Then we can integrate out degrees of freedom with energies of the order of the temperature to arrive at a new EFT (that we called pNRQCD\textsubscript{HTL}) in which the potential, among other Wilson coefficients, gets thermal corrections. In another case, if we consider that $T\sim \frac{1}{r}\gg m_D$, then we can start with NRQCD at $T=0$ and integrate at the same time the scales $\frac{1}{r}$ and $T$ to arrive at pNRQCD\textsubscript{HTL} in which now the potential has different thermal modifications.

An object that contains much information about the physics of quarkonium in a medium is the time-ordered propagator of the singlet. It encodes the medium modifications of the dispersion relation (corrections to the mass and the decay width), and its knowledge allows to obtain the spectral function (which is useful to compare with Lattice QCD results). Note that this object is not directly related to the number of quarkonium states in a plasma, so it is not possible to compute $R_{AA}$ only using it. Here we summarize essential points about our current knowledge about this object (based on pNRQCD computations in \cite{Brambilla:2008cx,Escobedo:2008sy,Escobedo:2010tu}):
\begin{itemize}
\item As can be seen looking at Eq. (\ref{eq:LanpNRQCD}), at $T=0$ its evolution can be described, at leading order, by a Schr\"{o}dinger equation. This is also true for the leading thermal corrections if $T\gg E$.
\item Thermal modifications of the potential induce an imaginary part which is related to the decay width of quarkonium. In the temperature regime $T\gg\frac{1}{r}\sim m_D$ the pNRQCD potential coincides with the one found in \cite{Laine:2006ns}. Note that the existence of this decay width does not imply that quarkonium states will disappear completely at large times. The square of the real-time propagator is not the probability of finding a singlet. It is the survival probability of a singlet without modifying the medium.
\end{itemize}
Until now we have discussed the application of EFTs ideas in the context of perturbation theory. However, it is possible to match NRQCD to pNRQCD in a completely non-perturbative way \cite{Brambilla:2004jw}. The pNRQCD potential is related to the expectation value of a Wilson loop, and it is possible to compute it using Lattice QCD. A recent non-perturbative evaluation can be found in \cite{Burnier:2016mxc}. These results show that the potential is complex and has a qualitative behavior similar to the one found in perturbation theory. The fact that both weak and strong coupling computations at finite temperature show that the potential is complex indicates that it is a quite general property and therefore it is crucial to understand what are the implications when computing $R_{AA}$. This issue is what we are going to discuss next.

\begin{figure}
\includegraphics[scale=0.5]{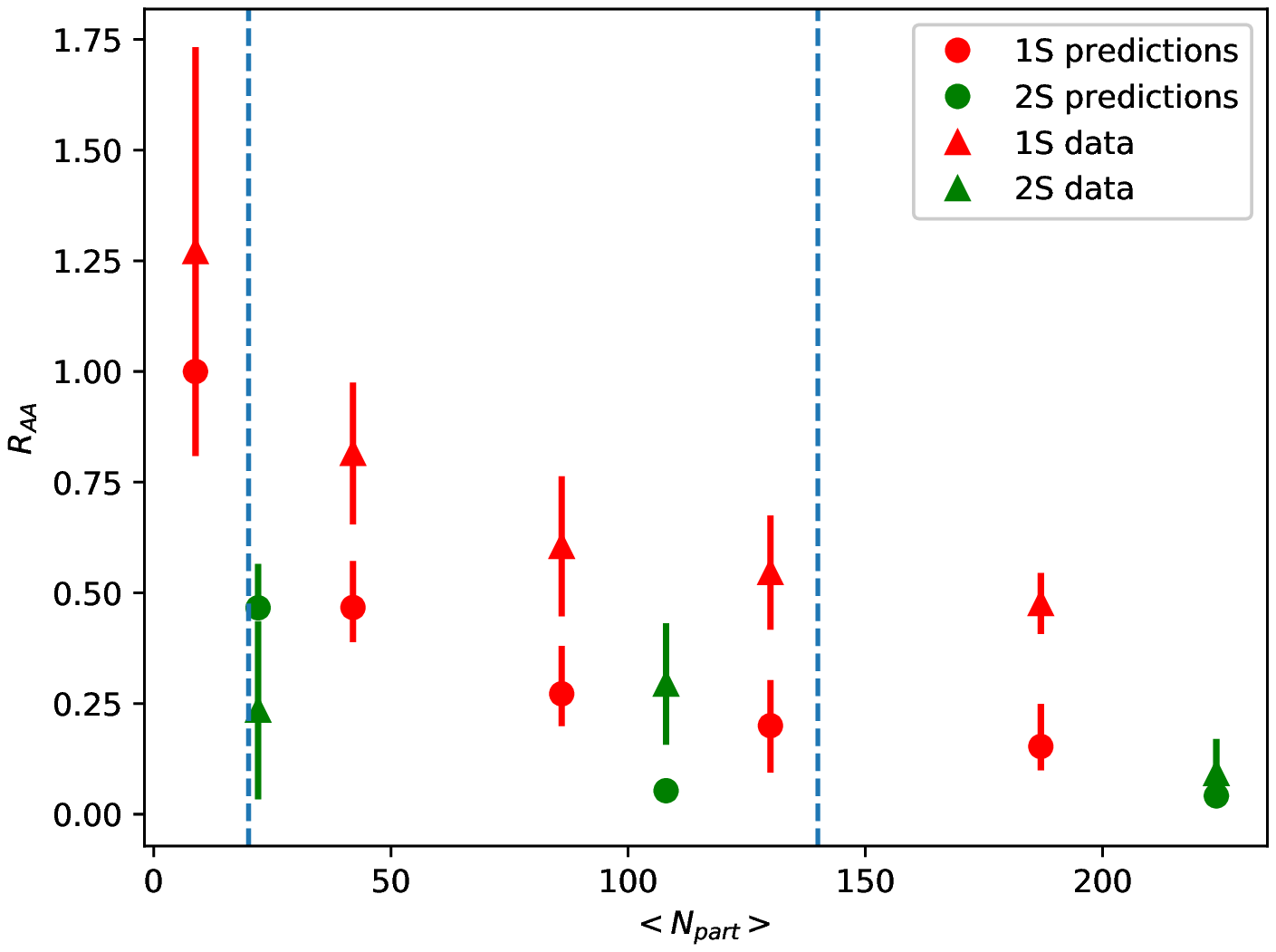}\includegraphics[scale=0.5]{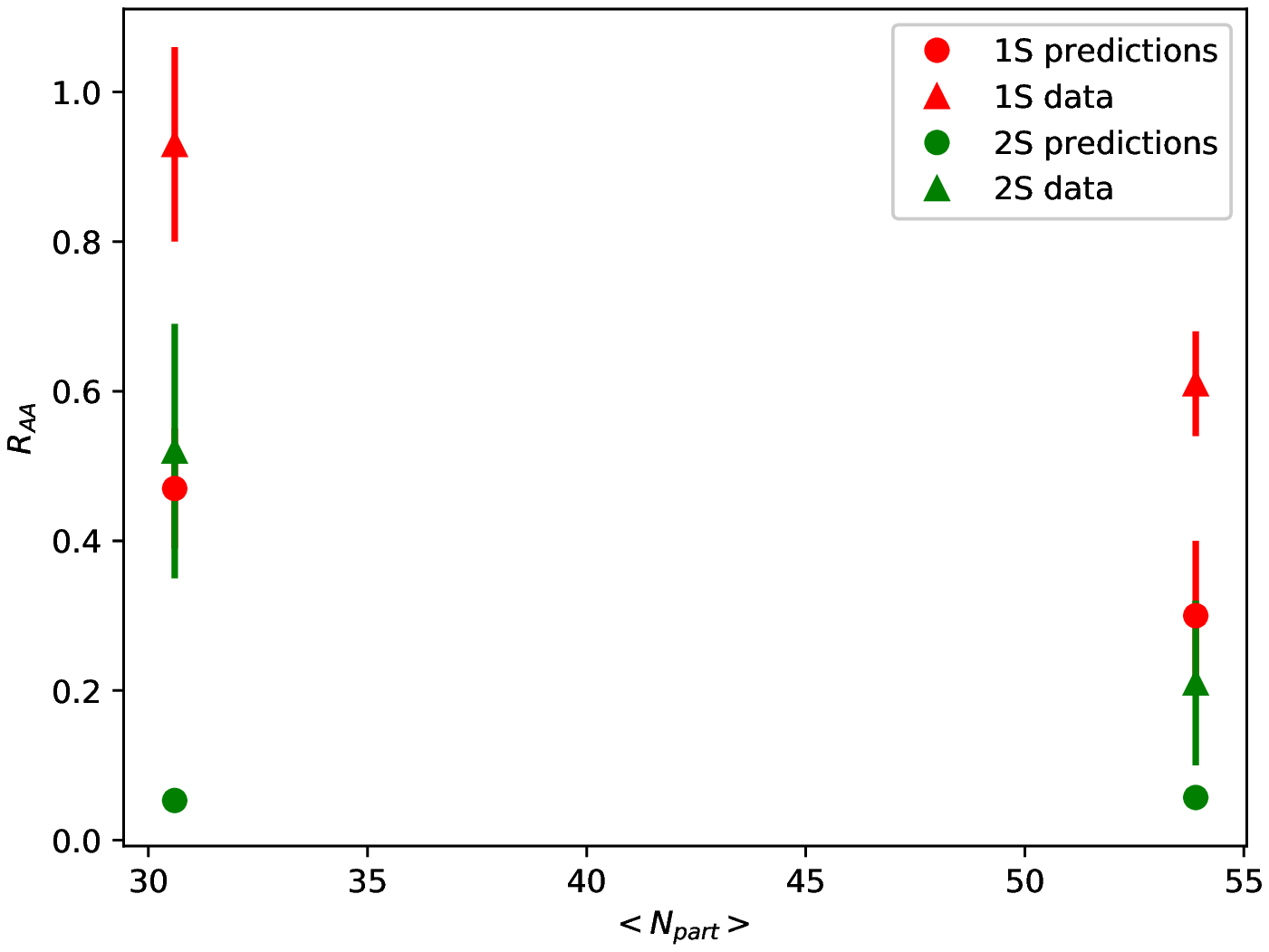}
\caption{Suppression factor of $\Upsilon(1S,2S)$ as a function of centrality (given by $\langle N_{part}\rangle$) in PbPb collisions. Left: comparison of our predictions with the experimental results of \cite{Khachatryan:2016xxp} at $\sqrt{s_{NN}}=2.76$~TeV. The vertical dashed lines highlight the window in which we expect our approximations to be valid. Right: the comparison of our prediction with the experimental results of \cite{Sirunyan:2018nsz} at $\sqrt{s_{NN}}=5.02$~TeV.}
\label{fig:RAA}
\end{figure}

The final goal is to understand quarkonium suppression in heavy-ion collisions, in other words, how the medium modifies the yield of bound states in nucleus-nucleus compared to pp collisions. In the pNRQCD framework, this is related to the quantity $\rho_{s}(x,y)=Tr(\rho S^{\dagger}(x)S(y))$ which is the projection of the density matrix to the subspace with one singlet. For example, the probability to find a $\Upsilon(1S)$ state is the projection of $\rho_s$ to the  $\Upsilon(1S)$ wave-function. The evolution of $\rho_s$ with time, under the assumption $\frac{1}{r}\gg T$, is described in \cite{Brambilla:2016wgg,Brambilla:2017zei}. The evolution of $\rho_s$ couples with that of the similar object $\rho_o$ (where o is for octet).  In the limit in which all thermal scales ($T$ and $m_D$) are bigger than the binding energy these coupled equations can be combined in a single Lindblad or GKSL equation \cite{Lindblad:1975ef,Gorini:1975nb}
\begin{equation}
\partial_{t}\rho=-i[H(\gamma),\rho]+\sum_{k}(C_{k}(\kappa)\rho C_{k}^{\dagger}(\kappa)-\frac{1}{2}\{C_{k}^{\dagger}(\kappa)C_{k}(\kappa),\rho\})\,.
\label{eq:Lindblad}
\end{equation}
In this equation, two non-perturbative parameters, that are in principle calculable using Lattice QCD, encode all the needed information about the medium. They are
\begin{equation}
\kappa=\frac{g^{2}}{6\,N_{c}}\,{\rm Re}\int_{-\infty}^{+\infty}ds\,\langle\,{\rm T}\,E^{a,i}(s,{\bf 0})E^{a,i}(0,{\bf 0})\rangle\,,
\end{equation}
and
\begin{equation}
\gamma=\frac{g^{2}}{6\,N_{c}}\,{\rm Im}\int_{-\infty}^{+\infty}ds\,\langle\,{\rm T}\,E^{a,i}(s,{\bf 0})E^{a,i}(0,{\bf 0})\rangle\,.
\end{equation}
To our knowledge, no Lattice computation on $\gamma$ is available. The results of the perturbative computation of the self-energy in \cite{Brambilla:2008cx} imply that $\gamma=-2\zeta(3)\,C_{F}\left(\frac{4}{3}N_{c}+n_{f}\right)\alpha_{s}^{2}(\mu_{T})\,T^{3}\,\approx-6.3\,T^{3}$. $\kappa$ is the heavy quark diffusion constant that has a very important role also in the physics of open heavy flavor. There are many Lattice QCD computations of this quantity in the literature. To our knowledge, the most recent one is \cite{Francis:2015daa}. In a future publication we will try to obtain $\kappa$ and $\gamma$ from lattice results on the mass shift and decay width of quarkonium \cite{kappa}. In fig. \ref{fig:RAA} we show the results of applying Eq. (\ref{eq:Lindblad}) together with the value of $\kappa$ of \cite{Francis:2015daa} and assuming $\gamma=0$ and we compare them with experimental results. The $\sqrt{s_{NN}}=5.02$~TeV results on the right-hand side of fig. \ref{fig:RAA} are new do not appear in previous publications.
\section{Jet quenching}
\label{sec:jet}
A useful tool to study jet related observables is Soft-Collinear Effective Theory (SCET) (see \cite{Fleming:2009fe} to find references to the original literature). SCET is typically described using light-cone coordinates, $x^\pm=\frac{x^0\pm x^3}{\sqrt{2}}$. It is suitable to describe processes in which particles with a high momentum $E$ in the + (or -) direction are involved. It is a theory equivalent to QCD when dealing with particles with virtuality smaller than $E^2$. Let us generically refer to $\lambda$ as the energy scale related with medium effects. Then SCET degrees of freedom can be divided into the following types:
\begin{itemize}
\item Collinear particles. They are particles with high energy but very small virtuality ($\lambda^{2}$). They are the main constituents of jets.
\item Hard-collinear particles. They are particles with high energy but not so small virtuality ($\lambda Q$).
\item Particles that form the medium, that at the same time can be divided in:
\begin{itemize}
\item Soft particles. They are such that when they hit a collinear particle, it becomes hard-collinear.
\item Glauber particles. They are similar to soft particles, but their momentum is such that when they hit a collinear particle, it keeps being collinear. Their crucial role when considering jets in a medium was highlighted in \cite{Idilbi:2008vm}.
\end{itemize}
\end{itemize}

A high energy particle undergoes two different type of processes when traversing a medium, quenching and broadening. By broadening we mean that the particle acquires a transverse momentum when going through the plasma. One can define a parameter, called $\hat{q}$, that encodes information about the strength of the broadening. We will define $\hat{q}$ in a more detailed way later. Quenching is the process in which a particle loses energy when traversing a medium. In all models of jet quenching, information about jet broadening is needed, and it happens that, in most of them, $\hat{q}$ contains all the needed information about the medium. In these proceedings, we are going to focus on the computation of jet broadening using SCET and Lattice QCD. There is also vast literature on the application of SCET to study directly jet quenching, which we are not going to cover. For more information about this, we refer to the many contributions on this topic that were also presented in hard probes 2018 \footnote{See the contributions of Ringer, Li, Chien, Yin and Kumar.}.

\begin{figure}
\includegraphics[scale=0.7]{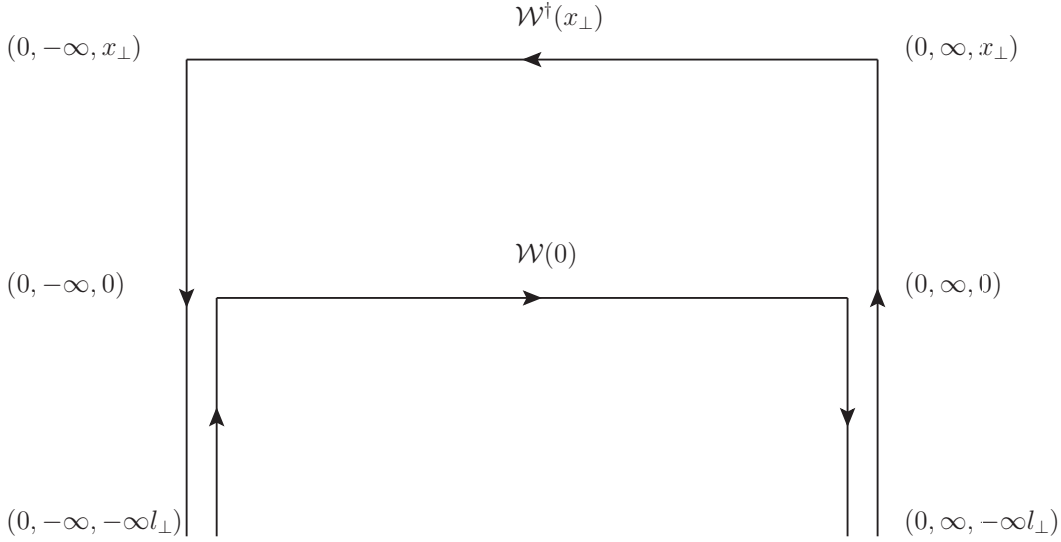}
\caption{ Wilson loop whose expectation value is equal to $P(x_\perp)$.}
\label{fig:Wloop}
\end{figure}

Now we discuss jet broadening and its relation with Lattice QCD. Let us use the following notation to write the momentum of a particle, $p=(p^+,p^-,p_\perp)$. Then, we can define the probability $P(p_\perp)$ that a high energy particle with an initial momentum $(0,E,0)$ turns into a particle with momentum $(\frac{p_\perp^2}{2E},E,p_\perp)$ after traversing a distance $L$ inside the medium. $\hat{q}$ can be defined as the second moment of this distribution
\begin{equation}
\hat{q}=\frac{1}{L}\int\frac{\,d^{2}k_{\perp}}{(2\pi)^{2}}k_{\perp}^{2}P(k_{\perp})\,.
\end{equation}
If the probability distribution is Fourier-transformed to coordinate space, then, it is equal to the expectation value of the Wilson loop along the light-cone direction shown in fig. \ref{fig:Wloop}. This relation was known before the application of SCET \cite{Baier:1996kr,Zakharov:1996fv,CasalderreySolana:2007pr} in a form that is valid in covariant gauges. The result was reobtained using SCET in \cite{D'Eramo:2010ak} and generalized to an arbitrary gauge in \cite{Benzke:2012sz}.
It is challenging to compute such an operator using Lattice QCD due to the Wilson lines going along the light-cone direction. However, it is possible to look at this Wilson line as connecting two points separated by a space-like distance but in a frame boosted with speed close to light \cite{CaronHuot:2008ni}.
This point of view allows using Electrostatic QCD \cite{Braaten:1995cm}, which is an EFT valid for static properties and energies below $\pi T$, to study the low energy contributions to $P(p_\perp)$, which are also the more non-perturbative ones. Several groups have used this combination of Lattice QCD with EFT ideas to obtain non-perturbative information about jet broadening \cite{Panero:2013pla,Laine:2013lia,Majumder:2012sh}.
\section{Conclusions}
\label{sec:concl}
The combination of EFTs with Lattice QCD computations is exceptionally effective to study hard probes in a medium. On the one hand, EFTs allow separating short-distance (high-energy) from long-distance (low-energy) physics. On the other hand, we can use Lattice QCD to obtain the needed non-perturbative information about long-distance effects.

We have illustrated this idea with the examples of heavy quarkonium and jet broadening. In the case of heavy quarkonium, we have shown that the definition of the potential and the applicability of potential models can be understood using pNRQCD. Lattice QCD provides information on a potential defined in this way, and it turns out to be complex. We can also compute $R_{AA}$ using the pNRQCD framework and, under some approximations,  two non-perturbative parameters, $\kappa$ and $\gamma$, contain all the needed information about the medium. In the case of jet broadening, the probability of acquiring a given transverse momentum is related to the expectation value of a Wilson loop. It is possible to compute the low energy contribution to this Wilson loop using Lattice QCD. This information then enters the predictions of jet quenching through the $\hat{q}$ parameter.
\section*{Acknowledgement}
The work of M. A. E has been supported by the Academy of Finland project 297058, by Ministerio de Ciencia e Innovacion of Spain under project FPA2017-83814-P and Maria de Maetzu Unit of Excellence MDM-2016- 0692; by Xunta de Galicia and FEDER.
\bibliographystyle{JHEP}
\bibliography{hp_escobedo}
\end{document}